\documentclass[11pt]{article}

\usepackage{fullpage}
\usepackage{amsmath}
\usepackage{amssymb}
\usepackage{mathrsfs}
\usepackage{setspace}
\usepackage{dsfont}
\usepackage{float}
\usepackage{subfigure}
\usepackage{adjustbox}
\usepackage{sidecap}
\usepackage{color,soul}
\usepackage{graphics}

\usepackage[latin1]{inputenc} 
\usepackage{color}
\usepackage{epstopdf}
\usepackage[american]{babel}
\usepackage{lineno}
\usepackage{setspace} 
\usepackage{cite}
\usepackage{authblk}
\usepackage[colorlinks=true]{hyperref} 
\hypersetup{
    bookmarks=true,         % show bookmarks bar?
    unicode=false,             % non-Latin characters 
    pdftoolbar=true,           % show Acrobat
    pdfmenubar=true,        % show Acrobat 
    pdffitwindow=false,      % window fit to page when opened
    pdfstartview={FitH},     % fits the width of the page to the window
    pdftitle={My title},         % title
    pdfauthor={Author},     % author
    pdfsubject={Subject},   % subject of the document
    pdfcreator={Creator},   % creator of the document
    pdfproducer={Producer},    % producer of the document
    pdfkeywords={keyword1} {key2} {key3},  % list of keywords
    pdfnewwindow=true,     % links in new window
    colorlinks=true,              % false: boxed links; true: colored links
    linkcolor=blue,               % color of internal links (change box color with linkbordercolor)
    citecolor=red,                % color of links to bibliography
    filecolor=magenta,         % color of file links
    urlcolor=cyan                 % color of external links
}

\newcommand{\be}{\begin{equation}}
\newcommand{\ee}{\end{equation}}
\newcommand{\bes}{\begin{equation*}}
\newcommand{\ees}{\end{equation*}}
\newcommand{\bea}{\begin{eqnarray}}
\newcommand{\eea}{\end{eqnarray}}
\newcommand{\beas}{\begin{eqnarray*}}
\newcommand{\eeas}{\end{eqnarray*}}

\newcommand{\bmat}{\begin{bmatrix}}
\newcommand{\emat}{\end{bmatrix}}

\def\CA{{\cal A}}
\def\CB{{\cal B}}

\title{A mathematical model for inter-specific interactions in seagrasses}

\author[1]{Eva Llabr{\'e}s}
\author[2]{Elvira Mayol}
\author[2]{N{\'u}ria Marb{\`a}}
\author[1]{Tom{\`a}s Sintes}
\affil[1]{Institute for Cross-Disciplinary Physics and Complex Systems, IFISC (CSIC-UIB),
Universitat de les Illes Balears, E-07122 Palma de Mallorca, Spain}
\affil[2]{Department of Global Change Research. Mediterranean Institute for Advanced Studies, IMEDEA (CSIC-UIB), C/ Miguel Marqu{\'e}s 21, 07190 Esporles (Mallorca), Spain}

\begin{document}

%\numberwithin{equation}{section}

\maketitle

\begin{abstract}
 Seagrasses are vital organisms in coastal waters, and the drastic demise of their population in the last decades has worrying implications for marine ecosystems. Spatial models for seagrass meadows provide a mathematical framework to study their dynamical processes and emergent collective behavior. These models are crucial to predict the response of seagrasses to different global warming scenarios, analyze the resilience of existing seagrass distributions, and optimize restoration strategies. In this article, we propose a model that includes interactions among different species based on the clonal growth of seagrasses. We present a theoretical analysis of the model considering the specific case of the seagrass-seaweed interaction between {\it Cymodocea nodosa} and {\it Caulerpa prolifera}. Our simulations successfully reproduce field observations of shoot densities in mixed meadows in the Ebro River Delta in the Mediterranean Sea. Besides, the proposed model allows us to investigate the possible underlying mechanisms that mediate the interaction among the two macrophytes.
\end{abstract}
\vspace{0.75cm}
\vspace{0.75cm}
\section*{INTRODUCTION}\label{sec:intro}

\par
Seagrass meadows are one of the most valuable structural elements in marine coastal areas since they provide ecosystemic services in the form of nutrient supply, refuge, and nursery ground to many species of fishes and invertebrates \cite{Costanza_1997}.  In addition to support marine biodiversity, they create architectural structure as benthic producers, contribute to the water quality and sediment stabilization \cite{Orth_2006}, protect coasts lines from strong waves \cite{FONSECA1992565,S_nchez_Gonz_lez_2011}, and are responsible for globally significant carbon sequestration \cite{bg-2-1-2005}. Moreover, seagrass beds support fisheries and provide livelihoods for millions of people in coastal communities \cite{Watson_1993}. Seagrass ecosystems, like many others, are under global threat due to anthropogenic impact \cite{Orth_2006,Hughes_2009}. Eutrophication, coastal development, water pollution, increased mooring activity, competition with invasive species, and global warming are some of the many causes that are leading seagrasses to experience an accelerating decline in the last decades \cite{Waycott:2009aa}. 
\vspace{0.6cm}
 \par
The critical demise of the worldwide population of seagrasses requires immediate and strategic actions for marine coastal protection. Mathematical models provide a theoretical framework to study the most relevant mechanisms that govern the dynamics of these ecosystems, and they can predict the possible future scenarios subjected to different environmental conditions. They can also evaluate the resilience of the existing population to stress factors and identify tipping  points, where a slight change in conditions may cause irreversible loss.  Thus, mathematical models constitute an essential tool to assess the health of ecosystems and generate more informed decisions for the sustainable management of marine coastal areas.  An agent-based spatial model for seagrasses and other organisms that exhibit clonal growth was proposed in \cite{TS2005}. This model collects detailed information of each shoot, and by iterating a set of empirically-based clonal growth rules, reproduces the non-linearities in the dynamics of a meadow  \cite{Sintes_2006}. The applications of the numerical simulations produced by the model are diverse, and range from assessing the CO$_2$ capture potential by seagrasses \cite{CDuarteCO2_2013} to estimate the age of living meadows \cite{Arnaud_Haond_2012}. A macroscopic description of the agent-based model was recently proposed to predict the spontaneous formation of spatial patterns in seagrass meadows at seascape level \cite{Ruiz-Reynes:2017aa}. In this model the change in the seagrass density obeys a set of coupled first order partial differential equations that include non-linear local and non-local interaction terms. By losing the detailed information on the shoot and ramet development, this macroscopic model works efficiently at larger spatial scales and it can be used to study the underlying mechanisms behind self-organization \cite{Ruiz_Reyn_s_2019, Ruiz_Reyn_s2_2020}.
\vspace{0.3cm}
 \par
Current models in the literature have not considered interactions among different species. Several interacting species commonly constitute seagrass ecosystems and form spatial distributions that range from perfectly separated domains to mixed meadows. The study of interspecific interactions is key to identify the main processes that shape the distribution of the seascape. For instance, the introduction of invasive species that compete for resources with native ones can drastically affect the habitat conditions \cite{norton1976sargassum,de_Vill_le_1995, Al_s_2016}. Also, the tropicalization of temperate latitudes due to global warming aggravates the spread of exotic species that are more resilient to higher water temperatures \cite{Rius_2014, Verg_s_2014,Wesselmann_2021}. Since each species have different responses to the thermal stress  \cite{Savva_2018, Collier_2011},  the addition of inter-specific interactions to the present numerical model is necessary to predict the dynamics of seagrasses under different global warming scenarios. 
\vspace{0.3cm}
 \par
In this work, we introduce a generalization of the agent-based model \cite{TS2005} that includes a cross-interaction term among clonal species either seagrasses or seaweeds. Such interaction is implemented in the local interaction term through coupling parameters that quantify the strength of the interaction between any pair of species. 
These parameters cannot be directly measured and need to be inferred indirectly by field observations. We will test the model and explore the role of the coupling for the specific case of the seagrass-seaweed interaction between {\it Cymodocea nodosa} and {\it Caulerpa prolifera}. These two species commonly form mixed meadows, and several studies showed that they negatively influence each other \cite{TUYA20131, PEREZRUZAFA2012101}. In this article, we propose a systematic way to fix the interaction parameters of the model and reproduce field observations of mixed meadows of {\it C. nodosa} and  {\it C. prolifera} in the Alfacs Bay (Ebro River Delta). 

\vspace{0.3cm}

\section*{MATERIALS AND METHODS}
	
\subsection*{Numerical Model}

\par
In this article we propose a numerical model to study the interactions among different species of seagrasses. Our model follows similar clonal growth rules as those formulated in  \cite{TS2005, Sintes_2006} for non-interacting seagrasses that we briefly summarized here. In their model, the development of clonal networks was simulated using a set of ecologically relevant parameters that can be easily derived from empirical observations, such as: the rhizome elongation rate $[v]$, that sets the horizontal spread of the clone; the branching rate $[\nu_0]$, that controls the capacity of the clone to form dense networks; the branching angle $[\phi]$, that determines the efficiency of the space occupation; the spacer length $[\delta]$, that measures the length of the piece of rhizome between consecutive shoots;  and the shoot mortality rate $[\mu]$ \cite{BELLTOMLIN,1bb9fcb800914db2af54226c5b443c5c}. The simulation starts placing a seed (a shoot carrying an apical meristem) and assigning to it a  unitary vector $\hat u$, randomly oriented, setting the direction of growth of the rhizome.  At each iteration, the following steps are repeated: 
\begin{enumerate}
 \item A rhizome, that originates in a randomly selected apex, is proposed to grow from its current position, $\vec{r}_0$, to $\vec{r}=\vec{r}_0+\delta \hat u$. The proposal is accepted if no other shoot is present within an exclusion area of radius $\sigma < \delta$ centered at $\vec{r}$. The value of $\sigma$ is set to avoid multiple shoots occupying the same position and to preserve the shoot density reported in natural stands of the species. Then, the apex is relocated at $\vec{r}$ where a new shoot will develop. In this process, the direction of growth, $\hat u$, does not change.
 \item Time is increased as $\Delta t= \delta / (v N_a(t))$, where $N_a(t)$ is the number of apices at time $t$. 
 \item A new branch, holding a growing apex, may develop at  $\vec{r}$ with a probability $p_{\nu}(t)=\nu_0  \Delta t  N_a(t)$.  The new branch will extend a long a new unitary vector $\hat u'$ forming an angle $\phi$ with $\hat u$ along the right or left side of $\hat u$, randomly chosen. Only one branch is possible at the position where the apex is located. 
 \item  Within this time step,  a number of shoots are removed from the meadow with a probability $p_{\mu}(t)=\mu \Delta t / N_s(t)$, where $N_s(t)$ is the number of shoots at time $t$. It is assumed the shoot mortality to be an age-independent event  \cite{Duarte_1994}.
\end{enumerate}
In this work, an alternative and more efficient type of exclusion procedure is implemented. A characteristic  value of the shoot density, according to the empirical observations, is set for each of the competing species, $\rho_{max}$. A square grid, representing the transects placed in the meadow, is superimposed on top of the continuum space. The grid spacing is set to $20 \, cm$. When a rhizome is proposed to extend into a cell which density has reached its saturation value  $\rho_{max}$, it will not advance in that iteration.
This method generates the same results as the exclusion area principle and reduces substantially the computing time. 
\vspace{0.3cm}
 \par
The application of the above mentioned growth rules lead to an homogeneous spatial development of the clones that strongly depend on the balance between the branching and the shoot mortality rates \cite{TS2005}, but is not able to reproduce the reported self-organized marine vegetation patterns, such as, stripes or fairy circles \cite{MMS12306,Borum2013, Frederiksen_2004, Pasqualini1999EnvironmentalII, Ruiz-Reynes:2017aa, van_der_Heide_2010}. This problem has been addressed assuming non-linear density-dependent growth rates that include facilitation and competitive interactions \cite{Ruiz-Reynes:2017aa}. Although the pattern formation in single specific meadows is directly related to the presence of non-local interactions, in the case of interacting species, and for the sake of simplicity, we will assume a local density dependent branching rate of the form: 
\be
\nu(\rho)=\nu_0+\alpha \hat\rho \left (1-\hat\rho \right ),
\label{singleeq}
\ee
where $\nu_0$ is the intrinsic branching rate that depends on external factors such as temperature or irradiance \cite{Duarte1989}. The local density dependence includes two terms: a facilitative one that is assumed to be linear and results from the positive contribution of the neighboring plants that dissipates the wave energy and prevents the shoot removal; and a non-linear competitive term that determines the environmental carrying capacity. The reduction in the branching rate may be the result of different mechanisms, such as competition for natural resources or self-shading in dense meadows \cite{Invers1997EffectsOP}. $\hat\rho = \rho/\rho_{max}$ is the normalized local density, and $\alpha$ is a coefficient that controls the strength of the interaction. Given the parabolic shape of the interaction, the growth of over- and under-populated areas is penalized, whereas regions around an optimal density  $\rho = \rho_{max}/2$ is favoured. 
The shoot mortality rate, $\mu$, is kept fixed. 
\vspace{0.3cm}
 \par
Equation \eqref{singleeq} can be easily extended to consider the interaction among $N$ species as follows.
We define a normalized local density for the species $i=1, \ldots, N$ as:
\be
\hat \rho_i={1\over \rho_{max,i}}{\left(\rho_i + \sum\limits_{j \ne i}\gamma_{ij} \rho_j\right)} \,,
\label{rhodef}
\ee
where $\rho_{max,i}$ is the saturation density for the $i-$species. $\gamma_{ij}$ is the coupling coefficient, which controls the strength of the competitive interaction between species $i$ and $j$. The higher their value is, the more one species is affected by the presence of the others, and vice-versa. The case $\gamma_{ij}=0$, reduces to the single-species case. 
This normalized density $\hat \rho_i$  is used in Equation \eqref{singleeq} to evaluate the branching rate for the $i-$species, $\nu_i(\rho_i)$. The form of the interaction term is chosen to be simple yet generic, and its biological interpretation will depend on the specific species we consider. Once the system is initialized with a random distribution of seeds of the competing species, the simulation proceeds as follows:
\begin{enumerate}
\item Since species have different characteristic growing times: $\tau _i={\delta_i/ v_i}$, at each iteration, one of the species is selected with probability  $p_i = \tau_i / \left ( \sum\limits_{i} \tau_i \right )$. 
\item The rhizome that originates in the $n^{th}$-apex of the $i$-species, randomly selected, is proposed to extend over a distance $\delta_i \hat u_i^{(n)} $.
\item The apex will be relocated to its new position and a new shoot will develop only if the normalized local density in the corresponding cell, given by the Equation \eqref{rhodef}, fullfils: $\hat \rho_i < 1$.
\item Time is increased by $\Delta t = \delta_i/(v_i N_a^T(t))$, where $N_a^T(t) $ is the total number of apices from all species at time $t$ 
\item A new branch with a growing apex will develop according to the branching rate  $\nu_i(\rho_i)=\nu_{0i}+\alpha_i \hat\rho_i (1-\hat\rho_i)$ with probability: $p_{\nu,i}(t)=\nu_i(\rho_i)  \Delta t  N_{a,i}(t)$, with $N_{a,i}(t)$ the number apices of the $i-$species.
\item During this time step, a number of shoots of the $i-$species are removed with probability $p_{\mu,i}(t)=\mu_i \Delta t / N_{s,i}(t)$, with $N_{s,i}(t)$ the number shoots of the $i-$species.
\end{enumerate}

The coupling coefficients $\gamma_{i,j}$ cannot be fixed by direct measurements and must be inferred indirectly by comparing the outcomes of the model to field observations. In this work, we will consider the seagrass-seaweed interaction between {\it C. nodosa} and {\it C. prolifera} and it will be used as a test case for the proposed model.
The set of parameters used as model inputs for each of the selected species is presented the Supplementary Materials A, according to averaged experimental observations. The saturation density for both species is selected to be $\rho_{max} = 1800$ shoots$/m^{2}$. In the Supplementary Materials C, we have studied the response of the interacting model proposed above for single species, taking into account local self-interaction term, \eqref{rhodef} with $\gamma_{ij}=0$. We find that the system tends to bi-stable density states that strongly depend on initial conditions. In the rest of this work, we will consider mortalities $\mu_i < \nu_{0,i}$. This condition ensures that the species are out of the bi-stability region. 
\vspace{0.3cm}
 \par
For two species ($N=2$), the number of coupling parameters reduces to $\gamma_{12}$ and $\gamma_{21}$. We will determine the values of these coefficients such that our simulations reproduce experimental observations.  In our simulations we have used a system size of $L \times L$,  with $L=  20 \, m $ and periodic boundary conditions. The results have been averaged over 15 independent realizations. 

\subsection*{Experimental design}

\par
In order to evaluate the coupling  parameters, the model outcome will be compared to the observed shoot densities in mixed meadows of  {\it C. nodosa} and {\it C. prolifera}. The field observation was conducted in the Alfacs Bay (Ebro River Delta), a shallow embayment (up to 6 m deep) with an estimated surface area of $50\, km^2$, located in the Northwestern Mediterranean. The fieldwork was performed on the northern shore of the embayment, which is covered by an extensive seagrass meadow of Cymodocea nodosa. The northern part of the bay receives seasonal freshwater inputs as runoff from rice paddy fields, with high nutrient and organic matter concentrations, and with suspended materials as well that increases water turbidity and, therefore, light deprivation, as compared to the southern part of the bay \cite{PEREZ1994249}. To avoid seasonal variability, plant sampling took place within a short  period of time  in two consecutive years (June 2018 and June 2019) when the seasonal peak in seagrass biomass and shoot density occurs \cite{2014ECSS..142...23M}. Points were randomly selected for a depth range, and distributed along, approximately, 10 km of seagrass meadow (See Supplementary Materials B for a map of the study site). At each sampling point, we collected with a hand-held corer (15 cm internal diameter) all the plant and algae biomass to a depth in the sediment of about 30 cm \cite{PerezMateoAlcoverroRomero}. Each sample was rinsed in situ thoroughly with seawater to eliminate the sediment, and organic material sealed in plastic bags, immediately frozen and carried to the laboratory where it was conserved at $-25^\circ$C until processing. Samples from 2018 were transported to the laboratory. 
Laboratory procedures involved separating plant and algae material of each sample into the different fractions (shoots, roots and rhizomes for   {\it Cymodocea nodosa}, and fronds, rhizoids and roots for   {\it Caulerpa prolifera}) and counting the shoots and fronds, respectively. In 2019, in order to optimize sampling effort, the shoot/frond density was counted directly on board, and no samples were transported to the laboratory.

  \vspace{0.4cm}

\section*{RESULTS}

\subsection*{Symmetric interaction between two species}
 
 \par
\begin{SCfigure}\centering
  \includegraphics[width=0.65\textwidth]{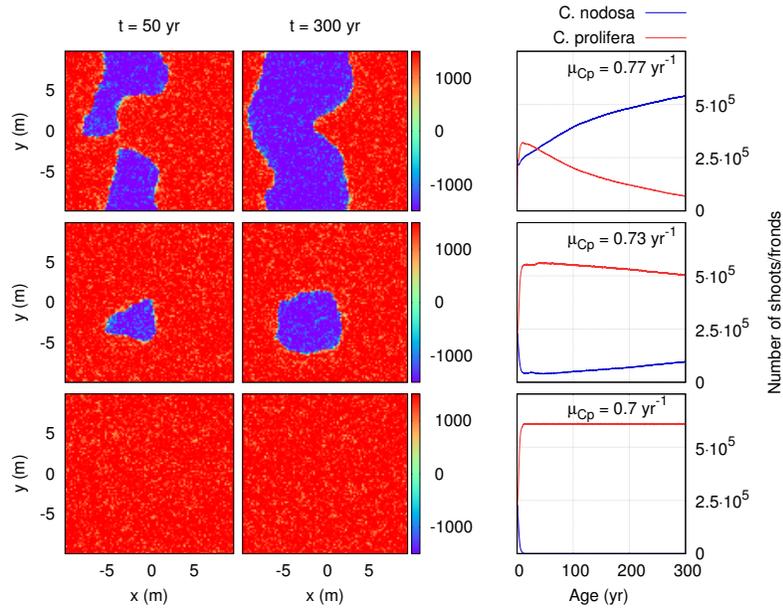}
    \caption{\label{fig:domain} \small Change in the spatial organization of {\it C. nodosa} and {\it C. prolifera} in a mixed meadow. The coupling coefficients are $\normalsize\boldsymbol{\gamma_{12}=\gamma_{21}=2}$.  The shoot mortality rate for {\it C. nodosa} is $\mu_{C_n}=0,92\, yr^{-1}$ , whereas the one for {\it C. prolifera} changes from $\mu_{C_p}=0.77$ to $0.70 \, yr^{-1}$, from top to bottom. Different snapshots are taken at $t=50,\,300\,yr$.  Regions colored in blue (red) are dominated by the presence of {\it C. nodosa} ({\it C. prolifera} ), and the color green represents regions of coexistence of both species. The color bar shows the difference between $\rho_{Cp}-\rho_{Cn}$, and it has units of shoots$/m^2$. In the right column, the  change in the average population of shoots for both species is shown.}
\end{SCfigure}
\begin{SCfigure}\centering
    \includegraphics[width=0.57\textwidth]{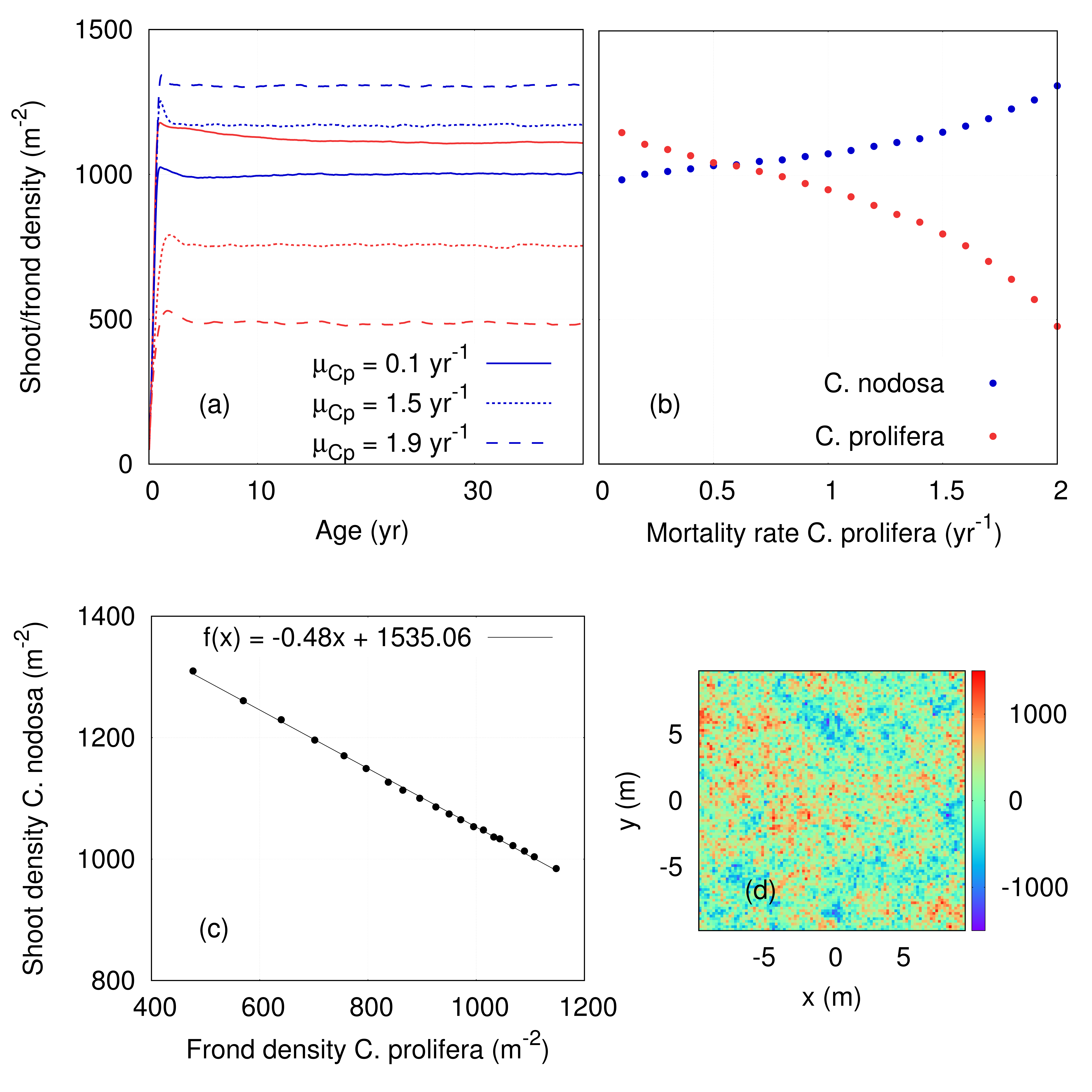}
    \caption{\label{fig:mixed05} \small  {Mixed meadows of \it C. nodosa} and  {\it C. prolifera} with a coupling coefficient $\normalsize\boldsymbol{\gamma_{12}=\gamma_{21}=0.5}$. (a) Change in the density profile for selected values of $\mu_{C_p}$. Red lines: data for {\it C. prolifera}; Blue lines: results for {\it C. nodosa}. (b) The saturation shoot densities of  both species vs  $\mu_{C_p}$. (c)  The shoot density of {\it C. nodosa} vs  {\it C. prolifera}  in mixed meadows for different values of $\mu_{C_p}$. (d) A snapshot of the meadow taken in the steady state regime at $t=10$ yr. for a {\it C. prolifera} mortality rate: $\mu_{C_p} = 0.65\, yr^{-1}$. The predominance of green color represents a mixed meadow solution. The color bar shows the difference between $\rho_{Cp}-\rho_{Cn}$, and it has units of shoots$/m^2$.}
 \end{SCfigure}
\begin{figure}[H]\centering
\includegraphics[width=1.0\textwidth]{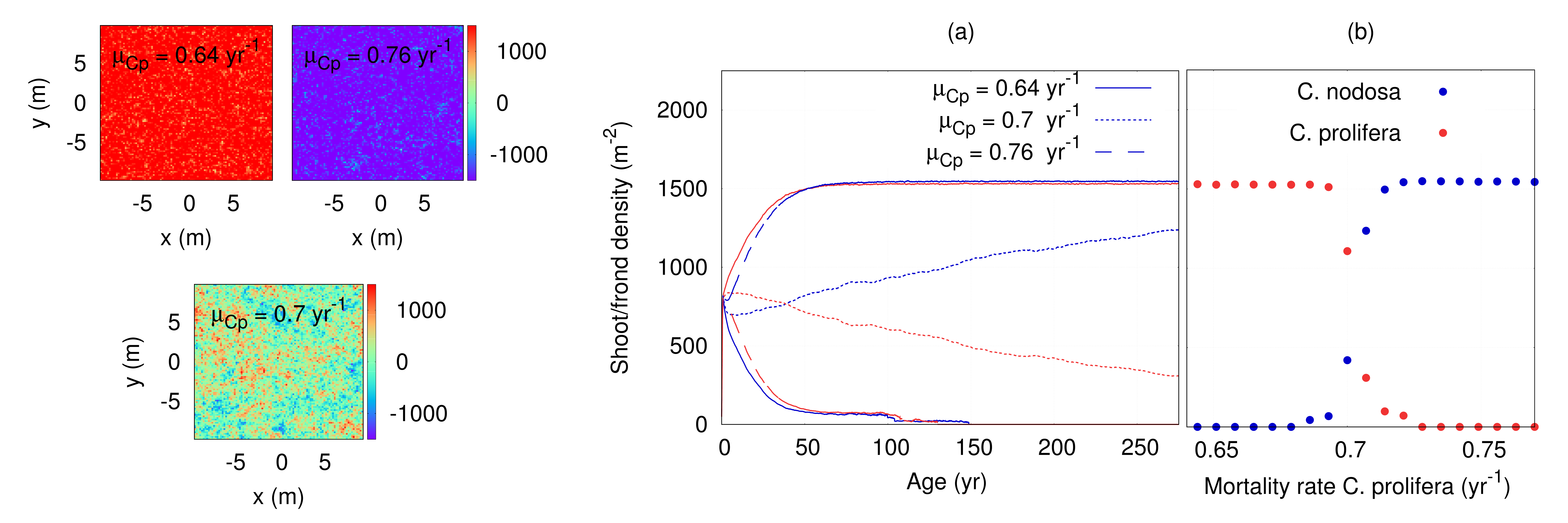}
\caption{\label{fig:domain1} \small{\it C. nodosa} and  {\it C. prolifera} meadows for coupling coefficients $\normalsize\boldsymbol{\gamma_{12}=\gamma_{21}=1}$. (Left) Snapshots of the three possible states:  mono-specific meadows of  {\it C. prolifera} in red  ($\mu_{Cp}=0.64\, yr^{-1}$) or  {\it C. nodosa} in blue ($\mu_{Cp}=0.76\, yr^{-1}$) (top row), and mixed meadows in green ($\mu_{Cp}=0.70\, yr^{-1}$) (bottom row). The color bar shows the difference between $\rho_{Cp}-\rho_{Cn}$, and it has units of shoot$/m^2$. (a) Change in the density profile of both species for selected values of $\mu_{Cp}$. Mixed meadows are found to be a transitory state. (b) Saturation shoot densities for both species vs. $\mu_{Cp}$.}
\end{figure}

We consider a scenario in which two species,  {\it C. nodosa} and {\it C. prolifera}, coexist in the same region. For simplicity, we start assuming symmetric couplings between them, i.e., $\gamma_{12}=\gamma_{21} =\gamma$.  
Our main result in this section is the identification of two distinct regions in the space of parameters:  $ \gamma > 1$, and $ \gamma < 1$. 
From the definition of the coupling coefficients in \eqref{rhodef}, we observe that $ \gamma > 1$ implies that the interaction between species is higher than the self-interaction. In this case, we expect meadows to separate into different domains to minimize the competition. The opposite happens for $\gamma < 1$, where the competition minimizes when the species mix. 
 In the following, we study the outcomes of our mathematical model in these different regimes.
 All the simulations have been initialized with a random distribution of seeds  that ensures a rather homogeneous spatial distribution. The initial shoot density for both species is $\rho_{0,_{Cn}} =\rho_{0,_{Cp}} = 25\, m^{-2}$, where the subindexes $Cn$ and $Cp$ stand for {\it C. nodosa} and {\it C. prolifera}, respectively.
 The  coefficient $\alpha$ has been set to $\alpha_{Cn}=\alpha_{Cp}=4$. 
 The value of the shoot mortality rate for {\it C. nodosa} is fixed to its average field observation value $\mu_{Cn}=0.92 \, yr^{-1}$ (Supplementary Materials A), whereas the one for {\it C. prolifera} has been varied in the region $\mu_{Cp} < \nu_{0,Cp}$ in order investigate the change in the population of shoots and their spatial distribution. 
 Here, we summarize the most relevant results of our simulations: 
 
   \vspace{0.4cm}
   
\begin{itemize}
 \item[-] {$\large\boldsymbol{\gamma > 1}$}: In this case, the competition between shoots of different species is strengthened, and it favors the spatial segregation of shoots in different domains. The results for $\gamma = 2$ are shown in Figure \ref{fig:domain}.  As expected, soon after the homogeneous initial condition has been set ($t< 5$ yr), well-defined domains separated by clear fronts emerge. For a shoot mortality rate for the {\it C. prolifera}, $\mu_{Cp} > 0.73\, yr^{-1}$, the area covered by {\it C. prolifera} shrinks, and the space left is occupied by {\it C. nodosa} that will become the dominant species. The situation reverts for $\mu_{Cp} < 0.7\, yr^{-1}$ in which {\it C. prolifera}  colonizes the whole meadow. The coexistence of different species is not possible and, after a transient period characterized by the spatial segregation of species, whose duration depends on $\mu_{Cp}$, the stable steady-state solution corresponds to a homogeneous mono-specific meadow. 

 \item[-] {$\large\boldsymbol{\gamma < 1}$}: The competition between species weakens, which favors the formation of mixed meadows. The results for $\gamma = 0.5$ are illustrated in Figure \ref{fig:mixed05}. The  steady-state behaviour is quickly achieved ($t < 2\, yr$) as seen in Figure \ref{fig:mixed05}(a). In this regime, the averaged shoot densities for the different species and different values of the shoot mortality rate of {\it C. prolifera}, $\mu_{Cp}$, are shown in Figure \ref{fig:mixed05}(b). Interestingly, the saturation densities for the {\it C. nodosa} vs.  {\it C. prolifera} in these mixed meadows follow a linear relationship (see Fig. \ref{fig:mixed05}(c)). The best fit to the data gives a slope of $-0.48\pm 0.01$. This result indicates that the occupation of {\it C. prolifera} fronds increases by a factor of two as the shoots of {\it C. nodosa} decreases. A representative snapshot of a mixed meadow after $t=10\, yr$  of growth at $\mu_{Cp}=0.65\, yr^{-1}$ is shown in Figure \ref{fig:mixed05}(d).

 \item[-] {$\large\boldsymbol{\gamma = 1}$}: This is the limit case between the two previous regions (Figure \ref{fig:domain1}). In this case, shoots of different species equally compete. The analysis of the saturation densities as a function of $\mu_{Cp}$ (Figure \ref{fig:domain1}(b)) shows that the coexistence region is extremely narrow in comparison with the one found for $\gamma = 0.5$ (Figure \ref{fig:mixed05}(b)). As a consequence, mixed meadows, that can last hundreds of years (see Figure \ref{fig:domain1}(a) for  $\mu_{Cp}=0.70\, yr^{-1}$), are a transient state towards a homogeneous mono-specific meadow. 

 \end{itemize}

 \subsection*{Experimental analysis and model validation}

\par
\begin{figure}[H]\centering
\includegraphics[width=1\textwidth]{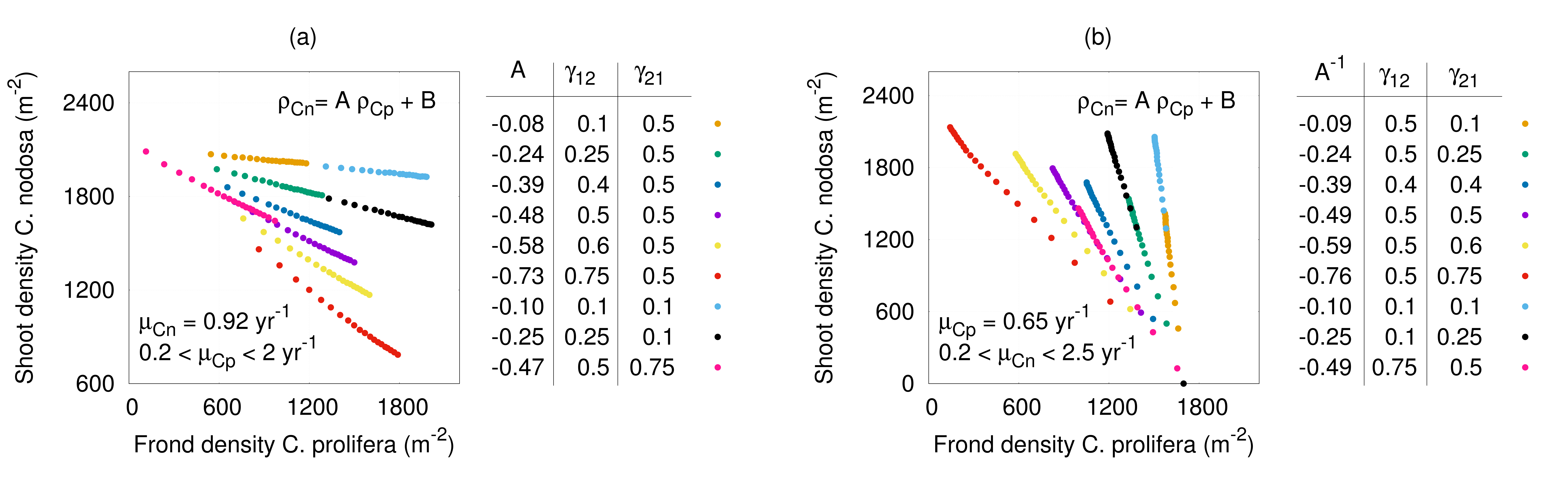}
\caption{\small \label{fig:slopeslope} Sensitivity analysis of the model for the coupling parameters:  $\gamma_{12},\,\gamma_{21}<1$.  We plot the shoot density of {\it C. nodosa} vs. {\it C. prolifera} in stable mixed meadows. Each color represents a different combination of the coupling coefficients. Each data point corresponds to a different value the {\it C. prolifera}  mortality rate,  $\mu_{Cp}$, in (a), and the {\it C. nodosa}  mortality rate, $\mu_{Cn}$, in (b).  The different sets of points fit the linear regressions of slope $A$, with an error not higher than $2\%$ in any of the cases.}
\end{figure}

In this section, we study the response of the model to non-symmetric coupling coefficients $\gamma_{12}\neq \gamma_{21} $.  One of our main observations is that the results from the symmetric case generalize and the formation of mixed meadows is guaranteed when the coupling coefficients are in the range $0<\gamma_{ij}<1$.
In  Fig. \ref{fig:slopeslope}, we plot the shoot densities of {\it C. nodosa} ($\rho_{Cn}$) vs. {\it C. prolifera} ($\rho_{Cp}$)  for different combinations of interaction parameters in $0<\gamma_{ij}<1$. 
We find that their steady-states are mixed meadow solutions that follow linear relations of the type $\rho_{Cn}=\CA \rho_{Cp}+\CB$, as in the symmetric case (Fig. \ref{fig:mixed05}(c)). 
In Fig. \ref{fig:slopeslope}(a), we kept the mortality of {\em C. nodosa} ($\mu_{Cn}$) fixed, while varying the mortality of {\em C. prolifera} ($\mu_{Cp}$). Interestingly, we observe that the coupling coefficient $\gamma_{12}$ governs the magnitude of the slope $\cal A$, which follows the relation $\gamma_{12}  \sim  -\cal A$. We could expect this relation from the linear stability analysis of the model at equilibrium (See Supplementary Materials D). For small values of $\gamma_{12}$, the density $\rho_{Cn}$ is less affected by a change in the mortality  $\mu_{Cp}$, resulting in a much smaller slope $\cal A$. Changing the other coefficient $\gamma_{21}$, translates the points along the same line since a decrease in $\gamma_{21}$ favors the growth of {\em C. prolifera} at the expense of {\em C. nodosa}, and vice-versa. An analogous analysis can be done for Fig. \ref{fig:slopeslope}(b), where we kept the mortality $\mu_{Cp}$ fixed and varied $\mu_{Cn}$.  Here,  the coefficient $\gamma_{21}$ controls the slope of the linear regression. In this case, it follows the inverse relation $\gamma_{21}  \sim  -{\cal A}^{-1}$,
and the coefficient  $\gamma_{12}$ does not alter the slope.  

\par
\begin{figure}[H]\centering
\subfigure{\includegraphics[width=0.75\textwidth]{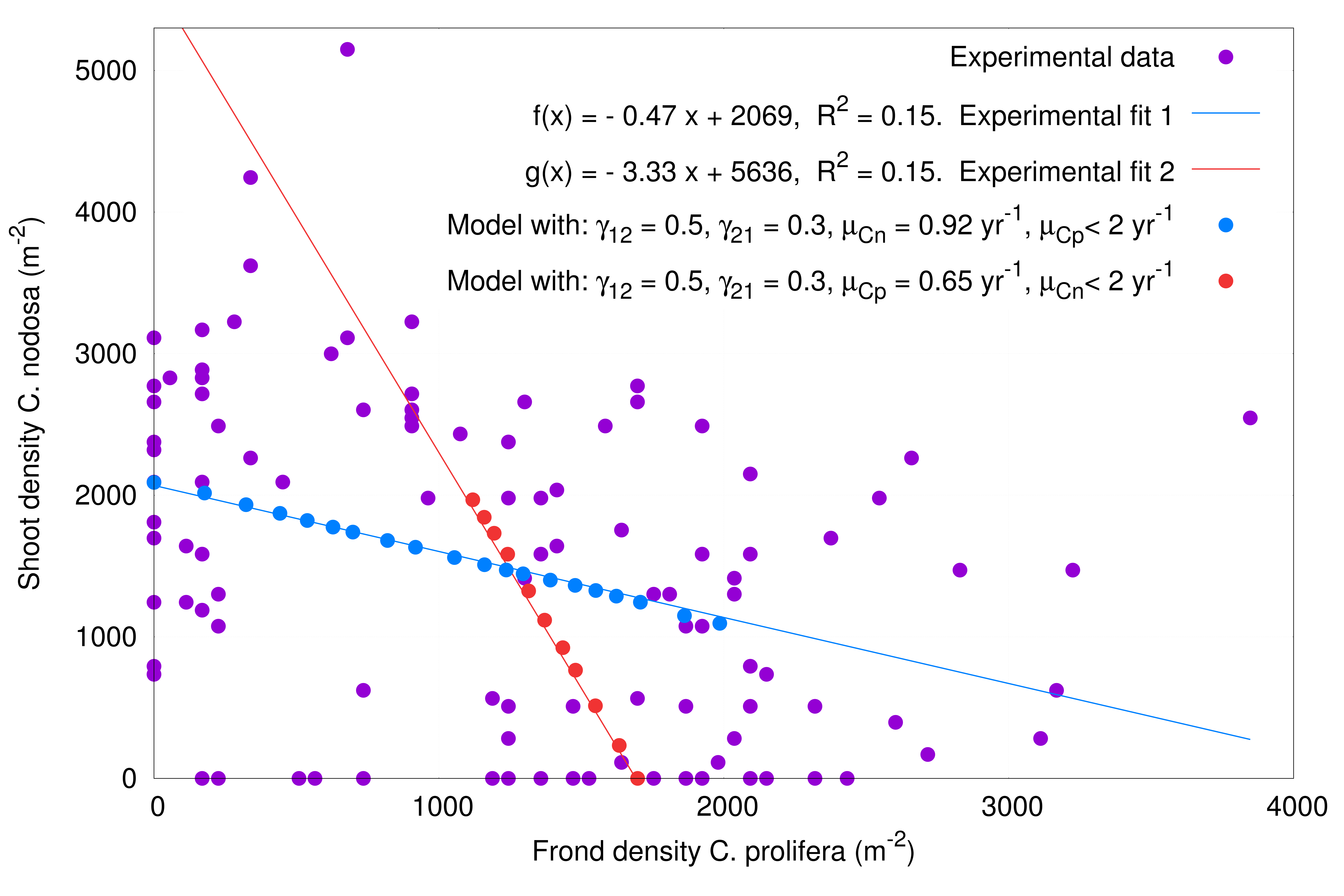}}
\caption{\label{fig:dades} \small {Comparison between the model outcome and the experimental data}. The purple dots correspond to experimental measurements of shoot densities of {\it C. nodosa} and {\it C. prolifera} in mixed meadows in the Alfacs Bay (Ebro River Delta,  Spain). The solid lines correspond to two types of linear fits of the experimental data, minimizing both the errors in the measurements of density of  {\it C. nodosa} (solid blue line), or minimizing the errors in  {\it C. prolifera} (solid red line). The dotted lines are the results of our simulations assuming coupling parameters $\normalsize\boldsymbol{\gamma_{12}=0.5$,\, $\gamma_{21}=0.3}$. The dotted blue line is reproduced by keeping the mortality of {\it C. nodosa} fixed, while varying the mortality {\it C. prolifera}, and vice-versa for the dotted red line. For a better comparison with the field measurements, the maximum shoot density has been adjusted to the average values found in the Alfacs Bay location ($\rho_{max,Cn}=2300\,m^{-2}$ and $\rho_{max,Cp}=1900\,m^{-2}$). }
\end{figure}

After the analysis of the model behavior, it is possible to determine the coupling coefficients $\gamma_{ij}$ in real meadows by comparing the results of our simulations with the experimental data. We measured the averaged shoot density in meadows of {\it C. nodosa} and {\it C. prolifera}  in 106 sampling points in the Alfacs Bay (Ebro River Delta - Spain), shown with purple dots in Figure \ref{fig:dades}. The presence of mixed meadows of these two species indicates values of the couplings $\gamma_{12}, \, \gamma_{21}<1$.  Although data is highly scattered, it is possible to find linear relationships between the shoot densities of {\it C. nodosa}, vs. {\it C. prolifera}:  $\rho_{Cn}=\CA \rho_{Cp}+\CB$. 
We use the least-square method to fit the data \cite{Watson_1967}. This method minimizes the errors of one of the data sets and considers the other as a control variable with no error. 
If we choose to minimize the errors in the measurements of $\rho_{Cn}$, the best fit to the data gives a slope of $\CA=-0.47 \pm 0.11$ (solid blue line).  This slope, within its error, is in a very good agreement with the model outcome for $\gamma_{12} \sim 0.5$  according to Fig. \ref{fig:slopeslope}(a). 
The linear regression that minimizes the errors in $\rho_{Cp}$ has a slope of $\CA=- 3.3 \pm 0.7 $ (solid red line), which coincides with the simulations when  $\gamma_{21}\sim 0.3$ (Fig.   \ref{fig:slopeslope}(b)).
Therefore, the numerical model nicely reproduces field data for $\gamma_{12}= 0.5$ and $\gamma_{21}= 0.3$. For these coupling coefficients, the simulations with fixed $\mu_{Cn}$ and varying $\mu_{Cp}$ generate the dotted blue line, and similarly for the dotted red line, by keeping $\mu_{Cp}$ constant and mo $\mu_{Cn}$. 
The match between model and data is one of our main results, which establishes $\gamma_{ij}$ as the key parameter to allow the connection between the numerical model and the experimental data and provide information on the strength of the interaction between species.

\par
\begin{figure}[H]\centering
\includegraphics[width=0.8\textwidth]{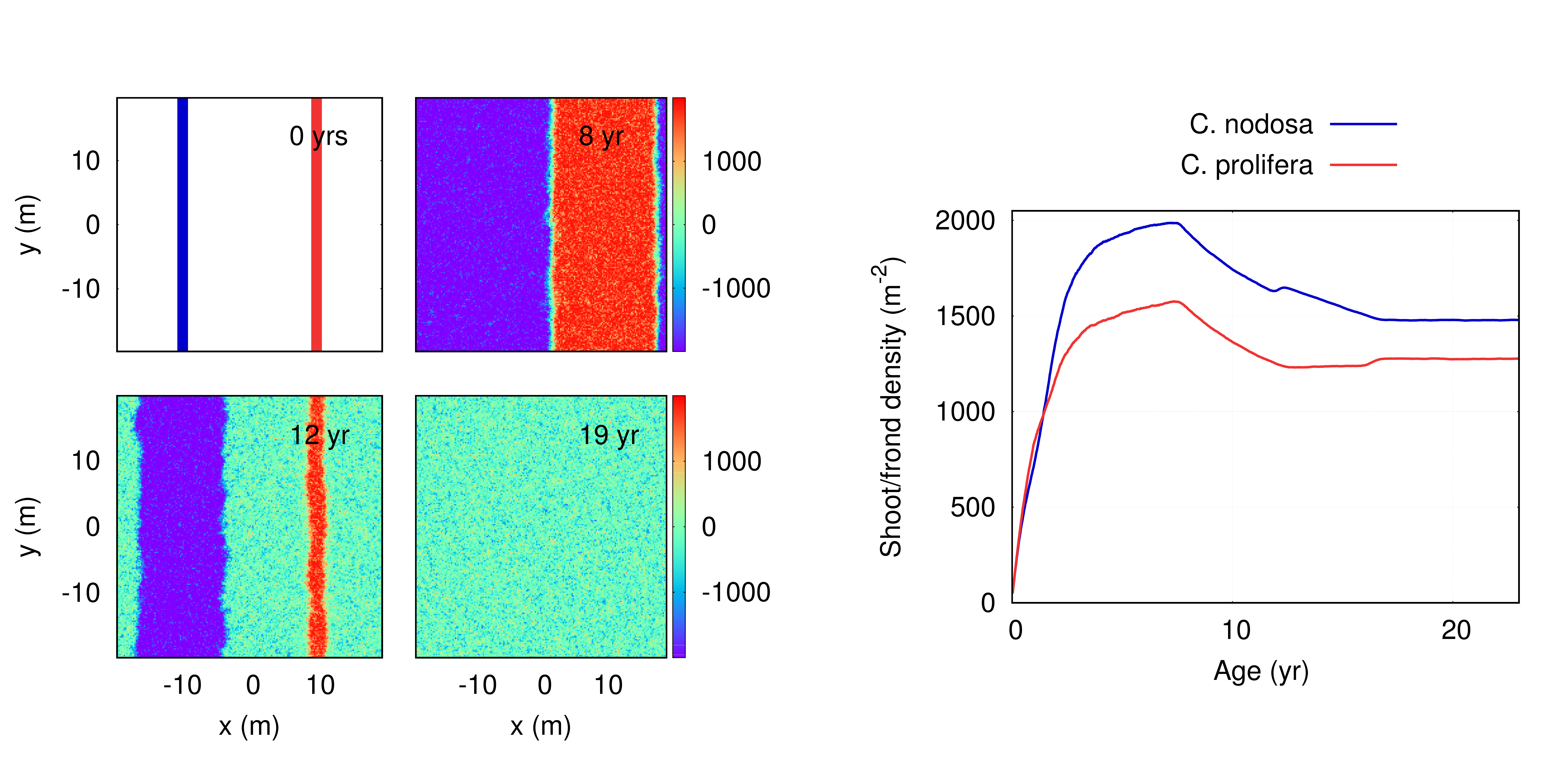}
 \caption{\label{fig:ci} \small (Left) Spatial distribution, at different time steps, of two competing species: {\it C. nodosa} (blue) and  {\it C. prolifera} (red), starting from two well defined striped patches separated from each other. As time evolves, both species cover all available space forming a mixed meadow (green). The color bar shows the difference between $\rho_{Cp}-\rho_{Cn}$, and it has units of shoots$/m^2$. (Right) Evolution of the averaged density of both species. The interaction coefficients are set to $  \gamma_{12}=0.5$ and $\gamma_{21}= 0.3$, and the mortality rates, $\mu_{Cp} = 0.65\, yr^{-1}$, and $\mu_{Cn} = 0.92\, yr^{-1}$.}
 \end{figure}

In all previous studies we assumed an initial condition consisting of a mixed meadow with the same amount of seeds of both species homogeneously distributed all over the available space. We analyze the sensitivity of the results to the initial condition. We will assume the case in which patches of different species are initially placed apart from each other. In a square region of size $L^2=40\times40\,m^2$, we locate a homogeneous density of seeds along two vertical stripes, $1\,m$  wide, one of {\it C. nodosa}, centered at $x= -10\,m$, and another of {\it C. prolifera}, centered at $x= 10\,m$ (Figure \ref{fig:ci} ($t= 0\,yr$)). During a short period of time, two separated mono-specific meadows develop. Shortly after that, the two domains  collide forming a clear front (Figure \ref{fig:ci} ($t \sim 8 \,yr$)). At this stage, the two species have reached their maximum shoot density values. This front quickly dilutes giving rise to a mixed meadow of {\it C. prolifera} and  {\it C. nodosa} (Figure \ref{fig:ci} ($t \sim 19 \,yr$)). The final state of the system is comparable to that collected assuming a homogeneous initial distribution of seeds (see Figure \ref{fig:mixed05}(d)).

\section*{DISCUSSION}

\par
In this paper we present the results of  an agent-based model that investigates the dynamics of growth of clonal organisms (like seagrasses and seaweeds) in a scenario of species competition. The interaction among species has been implemented through a coupling parameter, $\gamma_{ij}$, placed in the local interaction term (eq. \eqref{rhodef}) that quantifies the strength of the interaction between any pair of species. We found the value of this coupling parameter to determine the model outcome that ranges from separated and well defined mono-specific domains to mixed ones. A comparison between the model outcome and field measurements, conducted in the Alfacs Bay where large meadows  of {\it C. nodosa} and  {\it C. prolifera} are present,  allowed us to determine the coupling coefficients between these two species ($\gamma_{12}=0.5$, $\gamma_{21}= 0.3$) and it is one of the main findings of this paper (Fig. \ref{fig:dades}).  In our simulations, the choice of coupling parameters $\gamma_{ij}$ is decisive since they reflect the type and magnitude of the interaction between species:  $\gamma_{12}$ controls the influence of {\it C. prolifera} over {\it C. nodosa}, and vice-versa for $\gamma_{21}$. The asymmetry found in the results is a clear indication that the negative effect of the presence of  {\it C. prolifera} over {\it C. nodosa} is higher than vice-versa. The fact that both coefficients are in the $\gamma_{ij}<1$ regime implies  that the seagrass/seaweed competes less with the shoots of the other species than with their own. Therefore, in order to minimize the competition, species self-organize and form mixed meadows. This effect should typically be explained by several biologically relevant mechanisms related to the anatomy, metabolism, or other properties of the plants. Although {\it C. prolifera} and {\it C. nodosa} have comparable magnitudes in their clonal growth rates (Supplementary Materials A), they have very different root lengths: while {\it C. nodosa} extends into the sediment to a mean depth of $14.1 cm$, the roots of {\it C. prolifera} just elongate an average of $3 cm$  \cite{Duarte_1998,Bedinger_2013}. This characteristic allows the shoots of the two species to absorb the nutrients from different soil regions, making self-competition higher than the interspecific competition, favoring the appearance of mixed meadows. Therefore, our numerical analysis leads us to conjecture that the difference in root lengths between species is one of the main mechanisms that mediate the formation of mixed meadows of {\it C. nodosa} and {\it C. prolifera}. 
\vspace{0.3cm}
 \par
Our analysis is consistent with previous results found in the literature. In \cite{TUYA20131},  shoot densities in mixed meadows of {\it C. nodosa} and {\it C. prolifera} were collected in 16 sites across Gran Canaria island (Spain, Atlantic Ocean) in the summer season, when the meadows should be at their optimal growth. The best linear fit to their data ($ \rho_{1} =   {\cal A} \rho_{2} +  {\cal B}$), minimizing the errors in the measurements of {\it C. nodosa}, led to a slope $\CA=-0.71 \pm 0.35$ which is consistent with our findings (Fig. \ref{fig:dades}). Despite the similarities found, different slopes can be expected due to the very different conditions between the locations of both experiments: the sites in our experiment were located in the Ebro River Delta, that is characterized by very calm,  shallow, and  high-on-nutrient waters, while  sites in \cite{TUYA20131} were in Gran Canary island, whose coastline has a lack of pronounced geographical accidents and it is very exposed to winds and currents.
\vspace{0.3cm}
 \par
In \cite{TUYA20131}, they also performed an experiment to establish the direct effect of {\it C. prolifera} over {\it C. nodosa}. In mixed meadows, $100\%$ of the fronds of {\it C. prolifera} were removed. Eight months later, the density of shoots of {\it C. nodosa} increased $1.5$-times. This result is also consistent with our findings. When the two species are allowed to grow separately, setting appropiate initial conditions as those depicted in Fig. \ref{fig:ci}, {\it C. nodosa} patches reach a maximum density of $\rho_{Cn}=2000\,m^{-2}$ 
(maximum of the solid blue line at $t \sim 8 \,yr$). Once the two species meet and interact a stable mixed meadow of {\it C. prolifera} and {\it C. nodosa} develop with densities $\rho_{Cn}=1480\,m^{-2}$ and $\rho_{Cp}=1270 \,m^{-2}$. Thus, isolated meadows of {\it C. nodosa} have an average shoot density $1.35$-times higher that in mixed meadows. Therefore, our model supports the observations made in \cite{TUYA20131}, where they conclude that the appearance of {\it C. prolifera} partially contributes to the demise of the population of  {\it C. nodosa}. 
Also in \cite{PEREZRUZAFA2012101}, the changes of seagrass beds in Mar Menor Coastal Lagoon (Spain, Mediterranean Sea) were analyzed. They observed a decrease and almost total loss of  {\it C. nodosa} in deeper areas of the lagoon ($2-7m$) after a colonization event of {\it C. prolifera} in the early 1970's. Our simulations sustain the hypothesis that the invasion of {\it C. prolifera} should negatively influence the abundance of  {\it C. nodosa}. Still, this cannot solely explain the loss of {\it C. nodosa}, and we should consider other effects that deteriorate the seagrass. For example, light availability could be a key effect in the dynamics of \emph{C. nodosa} and \emph{C. prolifera} meadows, since it affects both species very differently. While \emph{C. prolifera} suffers from photoinhibition in transparent shallow areas, the shoot density of \emph{C. nodosa} could be limited by an increased content of silt, clay, organic matter, or phytoplankton\cite{PEREZRUZAFA2012101, P_rez_Ruzafa_2019, GARCIASANCHEZ201237}.
In \cite{BELANDO2021103415}, the most recent study of the distribution of {\it C. nodosa} and {\it C. prolifera} in the Mar Menor, they study the presence of mixed meadows in the deeper areas of the lagoon (less than $-4m$). Their data does not show a negative correlation between the biomass of both macrophytes, which is in disagreement with our findings in Fig. \ref{fig:dades}. The discrepancy between both observations might be related to the different environmental conditions among study sites, and it should be further investigated. 

In this work, we have presented the first numerical model that studies local interactions among different species of rhizomatous macrophytes. Although the present work has focused on mixed meadows of \emph{C. nodosa} and \emph{C. prolifera}, the proposed model can be applied to study the interactions between generic $N$-species. For example, by properly adjusting the clonal growth parameters and coupling coefficients, we could simulate the dynamics of tropical seagrass beds, with a large diversity of coexisting species \cite{SHORT20073}. Also, this numerical model can be used to predict the re-distribution of seagrasses over the ocean floor due to global warming, paying special attention to the competition between species with different thermal thresholds, such as \emph{P. oceanica} and  \emph{C. nodosa}, or the role of invasive species \cite{Savva_2018, Rius_2014}. Finally, this work sets the necessary framework to study long-range interactions between seagrass species. These types of interactions are often mediated by chemical concentrations and are responsible for the appearance of optimal sea-scape patterns \cite{Ruiz-Reynes:2017aa}. In future work, we will extend the current model to consider such effects.

\section*{ACKNOWLEDGMENTS}

We are especially grateful to Neus Sanmart{\'i}, Javier Romero, Marta P{\'e}rez, and Jordi Boada for their collaboration in the field observations and data collection.  
ELL and TS acknowledge the Research Grants: PRD2018/18-2 funded by LIET from the D. Gral. d'Innovaci{\'o} i Recerca (CAIB), RTI2018-095441-B-C22 funded by MCIN/AEI/10.13039/501100011033 and by European Regional Development Fund - "A way of making Europe", and MDM-2017-0711 funded by MCIN/AEI/10.13039/501100011033. 
NM and EM also acknowledge the Spanish Ministry of Science, Innovation and Universities (SuMaEco RTI2018-095441-B-C21).

\section*{AUTHORSHIP}

NM, EM, ELL, and TS conceived this study. ELL developed the models, produced the figures, and analyzed the data. EM carried out the field observations, and wrote the experimental methodology section. NM, EM, ELL, and TS analized the results. ELL wrote the initial draft and all coauthors contributed to editing the manuscript.

%\printbibliography
%\bibliographystyle{apalike}

%\bibliography{seagrasses}

\end{document}